\documentclass[doc,12pt]{apa6}\usepackage[]{graphicx}\usepackage[]{color}
\makeatletter
\def\maxwidth{ %
  \ifdim\Gin@nat@width>\linewidth
    \linewidth
  \else
    \Gin@nat@width
  \fi
}
\makeatother

\definecolor{fgcolor}{rgb}{0.345, 0.345, 0.345}

\usepackage{framed}
\makeatletter
 {\par\unskip\endMakeFramed%
 \at@end@of@kframe}
\makeatother

\definecolor{shadecolor}{rgb}{.97, .97, .97}
\definecolor{messagecolor}{rgb}{0, 0, 0}
\definecolor{warningcolor}{rgb}{1, 0, 1}
\definecolor{errorcolor}{rgb}{1, 0, 0}
\newenvironment{knitrout}{}{} 

\usepackage{alltt}
\usepackage[american]{babel}
\usepackage{csquotes}
\usepackage[style=apa,sortcites=true,sorting=nyt,backend=biber]{biblatex}
\DeclareLanguageMapping{american}{american-apa}
\usepackage{caption}
\usepackage[hidelinks=true]{hyperref}
\usepackage{bm}
\usepackage{mathtools}
\usepackage{dsfont}
\usepackage{amsfonts}
\usepackage{setspace}
\usepackage{tikz}
\usetikzlibrary{fit}
\usepackage{xspace}
\usepackage{listings}
\usepackage{dcolumn}
\newcolumntype{d}[1]{D{.}{.}{#1} }

\DeclareNameFormat{apaauthor}{%
  \ifthenelse{\value{listcount}=\maxprtauth\AND\value{listcount}<\value{listtotal}}
    {\addcomma\addspace\ldots\addspace}
    {\ifthenelse{\value{listcount}>\maxprtauth\AND\value{listcount}<\value{listtotal}}
      {}
      {\ifthenelse{\iffieldequalstr{doubtfulauthor}{true}}
        {\mkbibbrackets{\usebibmacro{name:apa:last-first}{#1}{#3}{#4}{#5}{#7}?}}
        {\usebibmacro{name:apa:last-first}{#1}{#3}{#4}{#5}{#7}}}}%
  \ifthenelse{\value{listcount}=\value{listtotal}}
    {\ifmorenames{\andothersdelim\bibstring{andothers}}{}}{}}

\newcommand{\proglang}[1] {\texttt{#1}\xspace}
\newcommand{\pkg}[1] {\texttt{#1}\xspace}
\newcommand{\code}[1] {\texttt{#1}\xspace}
\newcommand{\rpf} {\pkg{rpf}}
\newcommand{\R} {\proglang{R}}
\newcommand{\OpenMx} {\pkg{OpenMx}}
\newcommand{\RateMat} {\triangle\hat\theta\xspace}
\newcommand{\Prob} {\mathrm{Pr}}
\newcommand{\pick} {\text{pick}}
\newcommand{\E} {\mathbb{E}}

\newcommand{\logit} {\text{logit}}
\newcommand{\rpm}{\sbox0{$1$}\sbox2{$\scriptstyle\pm$} 
  \raise\dimexpr(\ht0-\ht2)/2\relax\box2 }

\title{A Computational Note on the Application of the Supplemented EM Algorithm to Item Response Models}

\shorttitle{Supplemented EM}

\author{Joshua N. Pritikin}
\affiliation{Virginia Commonwealth University}

\abstract{The EM algorithm is a method for
finding the maximum likelihood estimate of a model in the presence of missing data.
Unfortunately, EM does not produce a parameter covariance matrix for standard errors.
Supplemented EM \parencite[SEM;][]{meng1991} is one method for obtaining the parameter covariance matrix.
SEM is implemented in both
open-source \parencite[e.g.,][]{chalmers2012,pritikin2014a}
and commercial \parencite[e.g.,][]{ssi2013}
item response model estimation software.
However, the original formulation of SEM did not
adequately account for the limitations of IEEE~754 floating-point.
Agile-SEM, a novel refinement of SEM,
is proposed and compared against the original algorithm and
a recent refinement \parencite{tian2013}
in a variety of item response model simulation studies.
By controlling for the numerical noise intensity
on a per-parameter basis,
Agile-SEM demonstrated the best convergence properties,
accuracy, and efficiency while,
at the same time, requiring fewer tuning parameters.
Complete source code is made freely available.
The potential generalization of Agile-SEM
to other EM application besides item response models is
left as future work.}


\authornote{Joshua N.~Pritikin, Virginia Institute for Psychiatric and Behavioral Genetics, Virginia Commonwealth University.

Correspondence concerning this article should be addressed to Joshua N.~Pritikin,
Virginia Commonwealth University, 800 E Leigh St, Biotech One, Suite 1-133, Richmond, VA 23219.
E-mail: jpritikin@pobox.com}

\IfFileExists{upquote.sty}{\usepackage{upquote}}{}
\begin{document}

\maketitle

\section{Introduction}

Once a model is fit to data, it is routine practice to examine the degree of
confidence we ought to have in the parameter estimates.
This information is found in the parameter covariance matrix $V$,
and in summary form, as standard errors, $\sigma = \text{diag}(V)^\frac{1}{2}$.
The EM algorithm \parencite{dempster1977} is a method for
finding the maximum likelihood estimate (MLE, $\hat\theta$) of a model
in the presence of missing data.
For example, one EM algorithm of importance to psychologists and educators
is \textcite{bock1981} for implementation of Item Factor Analysis (IFA).
Unfortunately, the parameter covariance matrix is not an immediate output
of the EM algorithm.
Before exploring methods to obtain the parameter covariance matrix
in an EM context, the EM approach will be informally outlined.

Following traditional notation, let $Y_o$ be the observed data.
We want to find the MLE $\hat\theta$ of parameter vector $\theta$
for model $L(Y_o|\theta)$. Unfortunately, $L(Y_o|\theta)$ is
intractable or cumbersome to optimize.
The EM approach is to start with initial parameter vector $\theta^{t=0}$
and fill in missing data $Y_m$ as the expectation of $\{Y_m|Y_o,\theta^t\}$ (E-step).
In the case of \textcite{bock1981},
the missing data are the examinee latent scores (as determined by item parameters).
Together, the observed $Y_o$ and made-up data $Y_m$ constitute completed data $Y_c$.
With the parameter vector $\theta^t$ at iteration $t$,
we can use a complete data method to optimize $L(\theta|Y_c)$ and find $\theta^{t+1}$ (M-step).
With an improved parameter vector $\theta^{t+1}$,
the process is repeated until $\theta^{t} \approx \theta^{t+1} \approx \hat\theta$.
As a memory aid, the reader may prefer to associate the $m$ in $Y_m$ with \emph{made up} (not \emph{missing}).

In exponential family models,
the parameter covariance matrix $V$ is often estimated using the
observed information matrix.
The negative M-step Hessian
\begin{align}
\mathcal{I}(\hat\theta;Y_c) \approx -\frac{\partial^2 \log L(\theta|Y_c)}{\partial\theta\partial\theta} \label{eqn:info-complete}
\end{align}
is usually easy to evaluate but asymtotically underestimates the variability of $\mathcal{I}(\hat\theta;Y_c)$.
A better estimate is the negative Hessian of only the observed data $Y_o$,
\begin{align}
\mathcal{I}(\hat\theta;Y_o) \approx -\frac{\partial^2 \log L(\theta|Y_o)}{\partial\theta\partial\theta}.
 \label{eqn:ob-info}
\end{align}
Usually $\mathcal{I}(\hat\theta;Y_o)$ is difficult to evaluate;
One benefit of the EM method is the ability to optimize $L(\theta|Y_o)$
efficiently without evaluation of Equation~\ref{eqn:ob-info}.

To estimate the parameter covariance matrix in an EM context,
many methods have been proposed.
Some methods require problem specific apparatus such as
the covariance of the row-wise gradients \parencite{mislevy1984}
or a sandwich estimate \parencite[e.g.,][]{louis1982,yuan2013}.
For IFA models,
the Fisher information matrix can be computed analytically.
However, it requires a sum over all possible patterns \parencite{bock1981}.
Since such a sum is impractical for as few as 20 dichotomous items,
no further consideration of this method will be given.
Here we will focus on methods that are less reliant on problem specific apparatus.

Richardson extrapolation has been advocated \parencite{jamshidian2000}.
Central difference Richardson extrapolation evaluates the observed data
log-likelihood $\mathcal L(Y_o|\theta)$ at a grid of points
in the $\theta$ space to approximate the Hessian.
The distance between evaluations is controlled by a
perturbation parameter.
The perturbation distance is reduced on every iteration.
Precision is enhanced by extrapolating the change in
curvature between iterations.
Unfortunately,
the number of points required to approximate the Hessian is $1+r(N^2 + N)$
where $r$ is the number of iterations and $N$ is
the number of parameters in vector $\theta$ \parencite{gilbert2012}.
This limits the practical applicability of Richardson extrapolation
to models with a modest number of parameters.

We are aware of only two algorithms that (potentially) offer performance that scales
linearly with the number of parameters and require little
problem specific apparatus:
the direct method \parencite{oakes1999} and Supplemented EM \parencite[MR-SEM;][]{meng1991}.
MR-SEM grew to popularity in IFA software since,
at one time, MR-SEM was the more efficient in both accuracy and computation
time than other readily available methods \parencite{cai2008}.
Although the direct method is worthy of consideration,
MR-SEM has received more than twice as many citations.
Hence, we limit our focus to MR-SEM.

In acknowledgment that the potential efficiency of MR-SEM is poorly realized by
the original algorithm, a refinement was proposed \parencite[Tian-SEM;][]{tian2013}.
Tian-SEM was found to perform well in a comparison to other information matrix estimation methods in a simulation study
of unidimensional and multidimensional item response models \parencite{paek2014}.
However, some challenges arise when translating Supplemented EM family algorithms into a computer program.
To appreciate these challenges, it will be helpful to review some quirks that arise when
performing calculation with floating point numbers.

\subsection{IEEE 754 binary floating-point}

\begin{figure}
\def\svgwidth{5in}
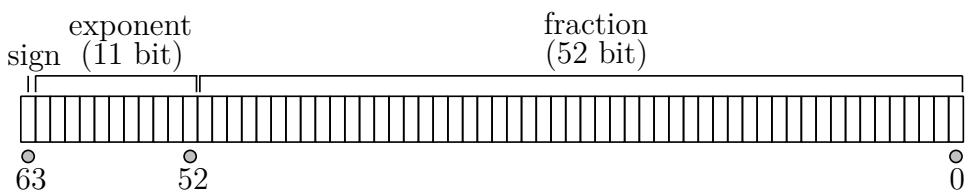
\caption{Binary layout of IEEE 754 double-precision floating-point.}
\label{fig:ieee754}
\end{figure}

A full length article is required to exhaustively detail the differences between
mathematically ideal real numbers and a floating-point
representation \parencite{goldberg1991}.
Here it is sufficient to observe some of the additive oddities
of floating-point.
Our examples will assume a double-precision (64-bit) representation,
but the essential arguments apply to any limited precision floating-point format.
In a floating-point representation,
the most significant digits of the number are stored in the \emph{fraction} part
and the magnitude is stored in the \emph{exponent} part (refer to Figure~\ref{fig:ieee754}).

A practical way to understand the implications of this format is
to consider,
\begin{align}
\arg\max_\lambda(|r|+\lambda) = |r| \label{eqn:fp-precision}
\end{align}
for a particular real number $r$.
That is, what is the largest $\lambda$ we can add to $|r|$
without changing $r$'s floating point representation?
For example, if we take $|r|=1$ then $\lambda\approx\exp(-37)$.
However, if we take $|r|=10^9$ then $\lambda\approx\exp(-17)$.
In other words, the magnitude of $r$ affects its precision.
Consider the convergence rule for an EM algorithm.
Convergence could be defined in terms of a norm of
the change in the parameter vector $||\theta^t - \theta^{t-1}||$
or in terms of the change in log-likelihood $|\mathcal L(Y_o|\theta^t) - \mathcal L(Y_o|\theta^{t-1})|$.
When the change in either quantity is less than some threshold then
the model can be declared converged.
However,
these tests are very different because the magnitude of the
log-likelihood is affected by the amount of data in the model.
If the same convergence threshold for log-likelihood was applied
uniformly then we would be implicitly requiring higher
convergence precision when there is more data, perhaps far in
excess of the parameter precision suggested by standard errors.

\subsection{Supplemented EM}

Supplemented EM (SEM) is based on
the observation that the information matrix of the completed data $\mathcal{I}(\hat\theta;Y_c)$
is the sum of the information matrices of the observed
$\mathcal{I}(\hat\theta;Y_o)$ and made-up data $\mathcal{I}(\hat\theta;Y_m)$  \parencite{orchard1972}.
With some algebraic manipulation we can rearrange the terms,
\begin{align}
\mathcal{I}(\hat\theta;Y_c) - \mathcal{I}(\hat\theta;Y_m) &= \mathcal{I}(\hat\theta;Y_o) \label{eqn:orchard} \\
\left[I - \underbrace{\mathcal{I}(\hat\theta;Y_m) \mathcal{I}^{-1}(\hat\theta;Y_c)}_{Y_m \text{ contribution}} \right] \mathcal{I}(\hat\theta;Y_c)
  &= \mathcal{I}(\hat\theta;Y_o). \label{eqn:sem-part1}
\end{align}
Intuitively, $\mathcal{I}(\hat\theta;Y_m) \mathcal{I}^{-1}(\hat\theta;Y_c)$
represents the fraction of information that $Y_m$ contributes to $Y_c$
in excess of $Y_o$ \parencite{dempster1977}.
One cycle of the EM algorithm can be regarded as a mapping $\theta \to
M(\theta)$. In this notation, the EM algorithm is
\begin{align}
\theta^{t+1} = M(\theta^t) \quad \mathrm{for}\ t \in \{0,1,\dots\}. \label{eqn:em-map}
\end{align}
If $\theta^t$ converges to some point $\hat\theta$ and $M(\theta)$ is
continuous then $\hat\theta$ must satisfy $\hat\theta \approx M(\hat\theta)$.
In the neighborhood of $\hat\theta$, by Taylor series expansion,
$\theta^{t+1} - \hat\theta \approx (\theta^t - \hat\theta)\RateMat$
where $\RateMat$ is the Jacobian of $M$ evaluated at the MLE $\hat\theta$,
\begin{align}
  \RateMat = \left.\frac{\partial M(\theta)}{\partial\theta}\right|_{\theta=\hat\theta}. \label{eqn:rate-matrix}
\end{align}
\textcite{dempster1977} showed that the rate of convergence is determined
by the fraction of information that $Y_m$ contributes to $Y_c$.
In particular, in the neighborhood of $\hat\theta$,
\begin{align}
\RateMat \approx \mathcal{I}(\hat\theta;Y_m) \mathcal{I}^{-1}(\hat\theta;Y_c). \label{eqn:sem-part2}
\end{align}
Combining Equations \ref{eqn:sem-part1} and \ref{eqn:sem-part2}, we obtain
$\mathcal{I}(\hat\theta;Y_o) \approx \left(I - \RateMat\right)\mathcal{I}(\hat\theta;Y_c)$.
Therefore, the inverse observed data parameter covariance matrix $V^{-1} \approx
\left(I - \RateMat\right)\mathcal{I}(\hat\theta;Y_c)$.

\subsection{SEM, from theory to practice}

The rate matrix $\RateMat$ from Equation~\ref{eqn:rate-matrix}
can be approximated using a forward difference method \parencite{meng1991}.
Let $d$ be the number of elements in vector $\theta$ so we can refer
to it as $\theta=\{\theta_1,\dots,\theta_d\}$.
Column~$j$ of $\RateMat$ is approximated by
\begin{align}
r_{.j}(\epsilon) = \frac{M(\hat\theta_1,\dots, \hat\theta_{i-1},
  \hat\theta_j + \epsilon, \hat\theta_{i+1}, \dots, \hat\theta_d) - M(\hat\theta)}{\epsilon}. \label{eqn:rij}
\end{align}
That is, we run 1 cycle of EM with $\theta$ set to the MLE
$\hat\theta$ except for the $j$th parameter of $\theta$ which is set to
$(\hat\theta_j + \epsilon)$ where $|\epsilon| > 0$.
(Note that indices $i$ and $j$ are interchangeable on the diagonal.)
Then we subtract $M(\hat\theta) \approx \hat\theta$ from the result and divide by the scalar $\epsilon$.
This amounts to numerically differentiating the EM map $M$.

Theoretically, accuracy improves as $\epsilon \to 0$.
In practice, however, this is arithmetic on a computer using a floating-point representation.
We cannot take $\epsilon \to 0$ but must pick a particular $|\epsilon|>0$.
The original formulation proposed to use the EM convergence history $\theta_j^t$
(where $\theta^t$ is the parameter vector $\theta$ at iteration $t$)
and compute the series of columns $\{ r_{.j}(\theta_j^t - \hat\theta_j)$,
$r_{.j}(\theta_j^{t+1} - \hat\theta_j), \dots \}$ until
$r_{.j}$ is ``stable'' from $t$ to $t+1$.
This procedure may initially seem appealing,
but note that the history of $\theta$ is a function of the starting
parameter vector $\theta^{t=0}$ and
no guidance was provided about appropriate starting values.
Regardless of starting values,
\textcite{meng1991} suggested that $r_{.j}$ could be declared stable
if no element changed by more than the square root of the
tolerance of an EM cycle.
For example, if the EM tolerance for absolute change in log-likelihood is $10^{-8}$
then the SEM tolerance would be $10^{-4}$.
Hence, the $j$th column of
$r_{.j}$ is converged when
\begin{align}
|r_{ij}(\theta_j^t - \hat\theta_j) - r_{ij}(\theta_j^{t+1} - \hat\theta_j)| < \mathrm{tolerance} \label{eqn:mr-criterion}
\quad \forall i \in \{1,\dots,d\}
\end{align}
But they remarked that the stopping criterion deserved further investigation.

With experience applying SEM to IFA models,
\textcite{tian2013} noted that parameter estimates $\theta^t$ typically
start far from the MLE $\hat\theta$ and approach closely only after a
number of EM cycles.
Starting SEM from $\theta^{t=0}$ is usually wasteful because
$r_{.j}$ does not stabilize until $\theta^t$ with $t$ close to
convergence.
During an EM run, the log-likelihood $\mathcal{L}$ typically changes
rapidly and then slowly as the parameter values are fine tuned.
They proposed $\delta^t = \exp\left(-\left| \mathcal{L}^t - \mathcal{L}^{t+1} \right|\right)$
as a ``standardized'' measure of closeness to convergence and suggested that
the best opportunity for SEM is history subset $\theta^t$ corresponding to
$\delta^t \in [.9,.999]$.
Unfortunately, in view of Equation~\ref{eqn:fp-precision},
$\delta^t$ is not a standardized metric
and works for models with approximately
the same amount of data as the models considered by \textcite{tian2013}.
More troubling,
\textcite{tian2013} did not address a weakness in
the original algorithm.
That is, MR-SEM provides no guarantee of convergence and frequently does not converge.
If a single parameter fails to converge then $\RateMat$ cannot be
estimated and all the extra computation is for naught.
In such an event,
\textcite{cai2012} suggested to lower the tolerance to
create a longer EM history.
Such a recommendation is not fatal but erodes confidence.
An analyst ideally wants reliable standard errors that are
unaffected by starting values or tolerance settings.

\begin{figure}
\begin{knitrout}
\definecolor{shadecolor}{rgb}{0.969, 0.969, 0.969}\color{fgcolor}
\includegraphics[width=\maxwidth]{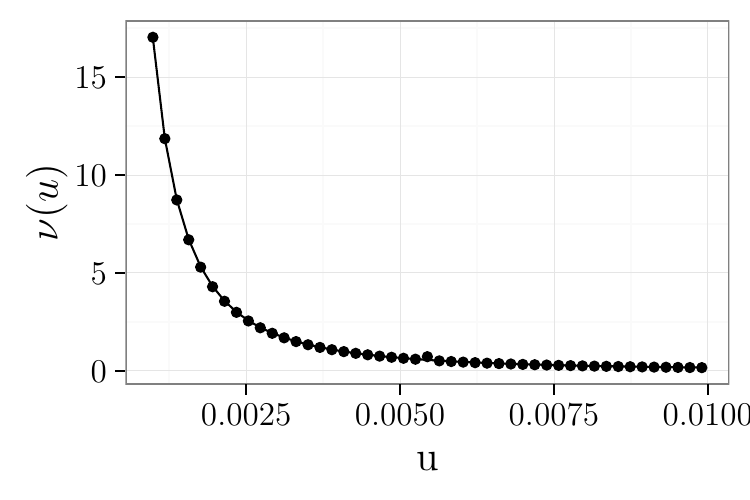} 

\end{knitrout}
\caption{Typical noise curve (Equation~\ref{eqn:asem-norm}) for a parameter of an IFA model.
The line is the model and the points are the measurements.
All points use $w=10^{-5}$.}
\label{fig:noise-curve}
\end{figure}

\begin{figure}
\begin{knitrout}
\definecolor{shadecolor}{rgb}{0.969, 0.969, 0.969}\color{fgcolor}
\includegraphics[width=\maxwidth]{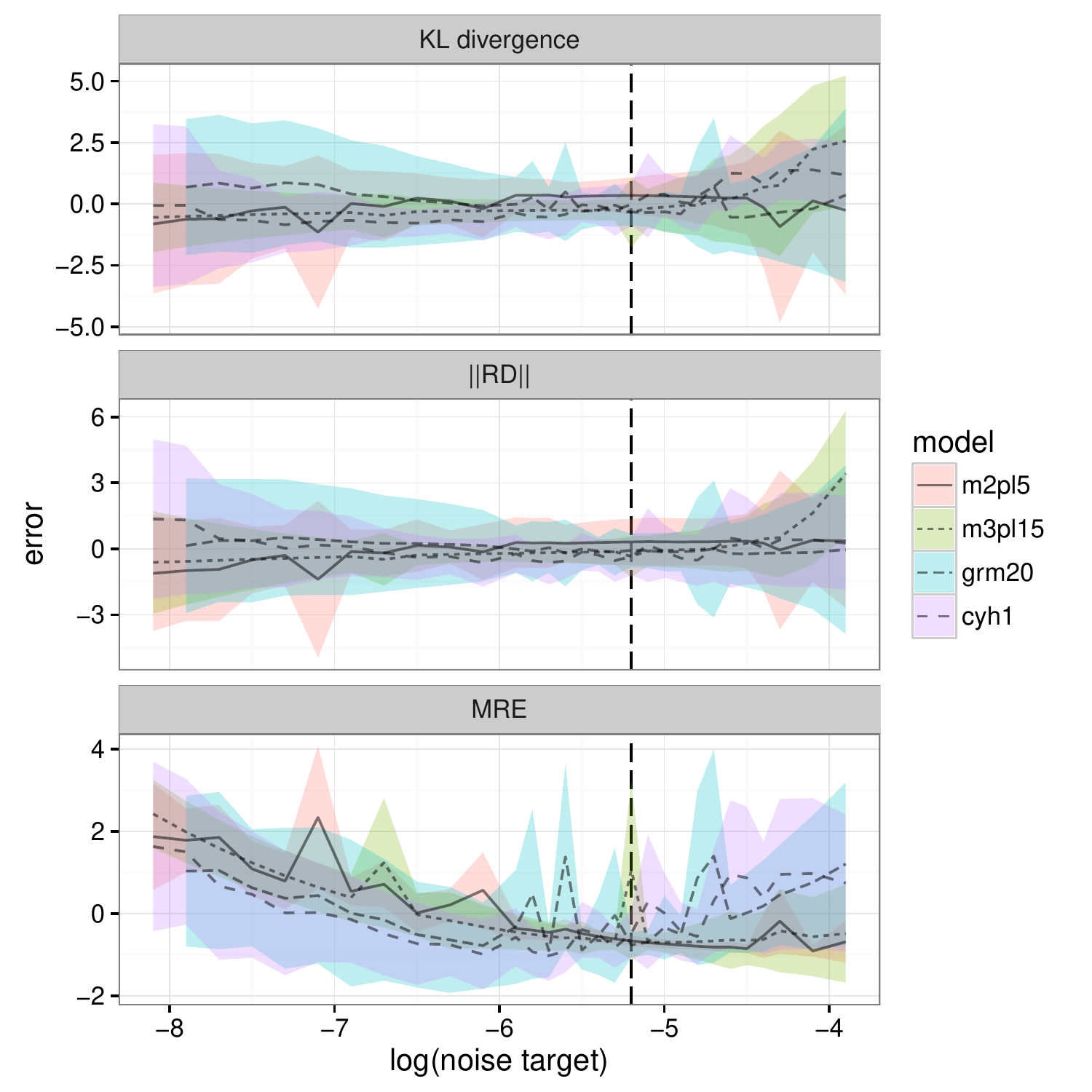} 

\end{knitrout}
\caption{Summary of parameter covariance matrix error (lower is better)
  for a range of noise targets over 100 Monte Carlo replications.
  The x axis shows the noise target but the x axis can also be regarded
  as a distance from the MLE $\hat\theta$ with -4 being closest to the MLE (with more numerical noise)
  and -8 furthest from the MLE (with less numerical noise but possibly a poor
  approximation of the gradient).
The selected noise target at $\exp(\ensuremath{-5.2})$ is shown with a vertical dashed line.
Outliers were defined as more than 10
median absolute deviation units from the median (1.11\% of the data).
Measurements were scaled to standard Normal after excluding outliers.
The shaded area shows the $\rpm 2 SD$ region.
Notice that the MRE measurements
corroborate the error intensity measured in the other 2 panels.
All three panels exhibit approximately the same U-shaped curve
with a minimum error and error variance between $-5.5$ and $-5$.
}
\label{fig:asem-target}
\end{figure}

Use of the EM convergence history may be counterproductive.
The definition of $\RateMat$ (Equation \ref{eqn:rate-matrix})
suggests that we should try to find
the smallest $|\epsilon|>0$
that produces a tolerable amount of numerical error.
It is easy to show that the magnitude of $||r_{.j}(\epsilon) - r_{.j}(\epsilon')||_1$
partially depends on the magnitude of $|\epsilon - \epsilon'|$.
This is a nuisance scaling factor.
Let us define a new norm that is the average
absolute difference between each pair of estimates divided by the spacing between probes $|w|>0$,
\begin{align}
\nu_j(u) = \frac{1}{d}\frac{\left|\left|r_{.j}(u-\frac{w}{2}) - r_{.j}(u+\frac{w}{2})\right|\right|_1}{w}. \label{eqn:asem-norm}
\end{align}
A noise curve is revealed if we plot $\nu_j$ with
an equal interval grid for $u$ (Figure~\ref{fig:noise-curve}).
A $w$ spacing of $10^{-5}$ is small enough that any change in $r_{.j}$ mostly
reflects change in noise intensity and not change in
the derivative.
Fortuitously, the noise curve is remarkably well modeled by
the regression formula
\begin{align}
\nu(u) = \frac{\beta}{u^2} + \text{error}. \label{eqn:asem-model}
\end{align}
If we fit measurements near the MLE for all parameters of Model grm20 (see Method),
the $R^2$ statistics are so close to 1 that
it is convenient to summarize the fit as
$\max\left[\log (1-R^2)\right]$=\ensuremath{-6.25}.

We can learn two things from the remarkable fit of the regression model.
Firstly, numerical noise $\nu(u)$ is closely related to how close
a probe is to the MLE and, secondly, the $\beta$ coefficient
is practically insensitive to $u$ near the MLE.
What does $\beta$ represent?
$\beta$ is a measure of signal strength and is the key to a fully automatic
version of SEM without tuning parameters.
For all the models examined in this article,
$\beta$ is on the order of $10^{-5}$ (range \ensuremath{9.99\times 10^{-6}} to \ensuremath{2.05\times 10^{-4}}).
Based on inspection of plots like Figure~\ref{fig:noise-curve},
we determined that a reasonable place to estimate $\beta$ is $u_1=10^{-3} + \frac{w}{2}$.
Since $\beta$ is fairly insensitive to the location where it is measured,
it seems unlikely that there is a substantially better place to measure $\beta$.
The scale of the parameters could affect the optimal $u_1$,
but fortunately, many popular IFA response models use parameters that
are roughly on the same scale (see Appendix~\ref{app:models}).

Remarkably, $\beta$ can help locate
where to approximate the Jacobian for a column of the rate matrix $\RateMat$ (Equation~\ref{eqn:rij}).
At this point, assume the coefficient $\beta$ is known.
If we neglect the error term in Equation~\ref{eqn:asem-model}
then we can solve $u_2$ for some target noise $\nu(u_2)$,
\begin{align}
u_2 = \left[\frac{\beta}{\nu(u_2)}\right]^{\frac{1}{2}}.
\end{align}
To determine a suitable target noise intensity $\nu(u_2)$,
a grid of candidate targets from $\exp(\ensuremath{-8.1})$ to $\exp(\ensuremath{-3.9})$
were tried (Figure~\ref{fig:asem-target}).
The definitions of these models and error quantities are given in the Method section.
A noise target of $\exp(\ensuremath{-5.2})$ was used in the reminder of this report.
It is not necessary to empirically evaluate $\nu(u_2)$.
We simply accept $r_{.j}(\epsilon=u_2)$ as the $j$th column of $\RateMat$.
We name this novel refinement of SEM \emph{Agile} because
$\epsilon$ is nimbly determined by an empirical noise measurement $\nu(u_1)$.
A pseudocode implementation is available in Appendix~\ref{app:pseudocode}.

\section{Method}

\subsection{Models}

We introduce a set of conditions designed to present a challenge to
parameter covariance matrix estimators.
We included underidentified models,
models with bounds,
and latent distribution parameters.
Underidentified models do not contain enough data
to uniquely identify the most likely model parameters.
The response probability functions employed
in the conditions are detailed in the \rpf package \parencite{pritikin2013a}
and also given in the Appendix~\ref{app:models}.
The structure of Models m2pl5, m3pl15, grm20, and cyh1
will be described.

Model m2pl5 contained 5 2PL items.
Slopes were 0.5, 1.4, 2.2, 3.1, and 4.
Intercepts were -1.5, -0.75, 0, 0.75, and 1.5.
Data were generated with a sample size of 1000 and all parameters were estimated.
Model m2pl5 is not always identified at this sample size.
This allowed us to examine the extent to which algorithms agreed
on whether a given model was identified or not.

Model m3pl15 contained 15 3PL items.
Slopes were set to 2 and items were divided into 3 groups of 5.
Each group had the intercepts set as in Model m2pl5 and
the lower bound parameters set to $\logit((1+g)^{-1})$ with $g$ as the group number (1-3).
A sample size of 250 was used.
For estimation, all slopes were equated to a single slope parameter.
To stabilize the model,
a Gaussian Bayesian prior on the lower bound (in logit units)
with a standard deviation of 0.5 was used \parencite[see][Appendix A]{cai2011}.

Model grm20 contained 20 graded response items with 3 outcomes.
Slopes were equally spaced from 0.5 to 4.
The first intercept was equally spaced from -1.5 to 1.5 every 5 items.
The second intercept was 0.1 less than the first intercept.
A sample size of 2000 was used and all parameters were estimated.
In the graded model, intercepts must be strictly ordered \parencite{samejima1969}.
The placement of intercepts so close together should
boost curvature in the information matrix.

\begin{table}[ht]
\centering
\begin{tabular}{rrrrrrr}
  \hline
Item & a1 & a2 & a3 & a4 & a5 & c \\ 
  \hline
1 & 1.00 & 0.80 &  &  &  & 1.00 \\ 
  2 & 1.40 & 1.50 &  &  &  & 0.25 \\ 
  3 & 1.70 & 1.20 &  &  &  & -0.25 \\ 
  4 & 2.00 & 1.00 &  &  &  & -1.00 \\ 
  5 & 1.40 &  & 1.00 &  &  & 1.00 \\ 
  6 & 1.70 &  & 0.80 &  &  & 0.25 \\ 
  7 & 2.00 &  & 1.50 &  &  & -0.25 \\ 
  8 & 1.00 &  & 1.20 &  &  & -1.00 \\ 
  9 & 1.70 &  &  & 1.20 &  & 1.00 \\ 
  10 & 2.00 &  &  & 1.00 &  & 0.25 \\ 
  11 & 1.00 &  &  & 0.80 &  & -0.25 \\ 
  12 & 1.40 &  &  & 1.50 &  & -1.00 \\ 
  13 & 2.00 &  &  &  & 1.50 & 1.00 \\ 
  14 & 1.00 &  &  &  & 1.20 & 0.25 \\ 
  15 & 1.40 &  &  &  & 1.00 & -0.25 \\ 
  16 & 1.70 &  &  &  & 0.80 & -1.00 \\ 
   \hline
\end{tabular}
\caption{Data generating parameters for Model cyh1.
Group 2 did not contain items 13-16. Nonzero parameters were estimated.} 
\label{tab:cyh1-item}
\end{table}

The first simulation study from \textcite{cai2011} was included.
Model cyh1 was a bifactor model with 2 groups of 1000 samples each.
Group 1 had 16 2PL items with the latent distribution fixed to standard Normal.
Group 2 had the first 12 of the items from Group 1.
All item parameters appearing in both groups were constrained equal.
Data generating parameters for the items are given in Table~\ref{tab:cyh1-item}.
The latent distribution of Group 2 was estimated.
Latent distribution generating parameters were 1, -0.5, 0, 0.5 and 0.8, 1.2, 1.5, 1,
for means and variances respectively.

In addition, a 20 item 2PL model and the model from the second
simulation study of \textcite{cai2011} were examined.
Little additional insight was gained from these models and we do not
report them here in detail.
However, this work indicated that our results
generalize to the nominal response model (see Appendix~\ref{app:models}).

All item response models used a multidimensional parameterization
(slope intercept form instead of discrimination difficulty).
Hence, intercepts were multiplied by slopes in Models m2pl5, m3pl15, and grm20.
Both the original formulation of Supplemented EM and \textcite{tian2013}
strongly depend on the parameter convergence trajectory.
Therefore, it is crucial to report optimization starting values.
In general, all slopes were started at 1, intercepts at 0, means at 0, and variances at~1.
For Model m3pl15, all lower bounds were started at their true value.
Since the intercepts of the graded model cannot be set equal,
for Model grm20, intercepts were started at 0.5 and -0.5 respectively.

\subsection{Monte Carlo estimates}

All models were subjected to 500 Monte Carlo trials
to obtain the ground truth for the parameter covariance matrix.
For each trial, data were generated with the \texttt{rpf.sample} function
from the \rpf package \parencite{pritikin2013a}.
Models were fit with \textcite{bock1981} as implemented in
the IFA module of \OpenMx with EM acceleration enabled \parencite{pritikin2014a,varadhan2008}.
For multidimensional models,
\textcite{cai2010} was used for analytic dimension reduction.
The EM and M-step tolerance for relative change in log-likelihood,
\begin{align}
\left| \frac{\mathcal{L}^t - \mathcal{L}^{t+1}}{\mathcal{L}^t} \right|,
\end{align}
were set to $10^{-9}$ and $10^{-12}$, respectively.
The use of relative change
removes the influence of the magnitude of $|\mathcal{L}|$ on the precision of $|\mathcal{L}|$.
In models where the latent distribution was fixed,
numerical integration was performed using a standard Normal prior.
Single dimensional models used an equal interval quadrature
of 49 points from Z score $-6$ to 6.
The multidimensional model used an equal interval quadrature
of 21 points from Z score $-5$ to 5.
The computer used was running GNU/Linux
with a 2.40GHz Intel~i7-3630QM~CPU and ample RAM.
Table~\ref{tab:mcData} summarizes the results.

\begin{table}[tbp]
\centering
\begin{tabular}{rrrrrrr}
  \hline
 & \#P & Unidentified & $\log(CondNum)$ & $\max(|bias|)$ & $||bias||_2$ & $\log(|V^{-1}|)$ \\ 
  \hline
m2pl5 & 10 & 13 & 16.1 & 0.665 & 0.84 & 35 \\ 
  m3pl15 & 31 & 6 & 8.5 & 0.306 & 0.55 & 90 \\ 
  grm20 & 60 & 0 & 16.1 & 0.111 & 0.22 & 369 \\ 
  cyh1 & 56 & 1 & 8.5 & 0.055 & 0.14 & 281 \\ 
   \hline
\end{tabular}
\caption{Descriptive summary of the Monte Carlo simulation studies. The first column is the number of free parameters in the model. Where the \emph{unidentified} column is 0, all trials were included. Trials were considered unidentified if the iteration limit was reached or the log condition number using the covariance of the gradients was greater than $\log(CondNum)$. $V$~is the Monte Carlo parameter covariance matrix.} 
\label{tab:mcData}
\end{table}

The condition number of the information matrix
is the maximum singular value divided by the minimum singular value and
provides a rough gauge of the stability
of a solution \parencite[][p.~239]{luenberger2008}.
For example, models that are amply overspecified have a condition number close to 0
whereas slightly overspecified models will have a large positive condition number.
When the information matrix is not positive definite then
the MLE is unstable and may be a saddle point \parencite[][p.~190]{luenberger2008}.
For reference, bias is defined as $\E\,\theta - \hat\theta$ (columns 4 and 5) and
the Monte Carlo parameter covariance matrix is simply
the covariance of each trial's MLE $\hat\theta$ as the rows of data (column 6).

\subsection{Measures of precision}

Kullback-Leibler~(KL) divergence was used to measure
the precision of a parameter covariance matrix estimate.
For a $0$ mean multivariate Normal distribution,
\[
D_{KL}(\Sigma_{true}, \Sigma) =
\frac{1}{2} \left[ Tr(\Sigma^{-1}\Sigma_{true}) - K -
\log\left( \frac{|\Sigma_{true}|}{|\Sigma|} \right) \right]
\]
where $K$ is the dimension of $\Sigma$.
KL divergence is a comprehensive quality metric,
but we may only be interested in the standard errors
on the diagonal.
The parameter variances could be
more accurately estimated than the covariances.
Therefore, a metric based only on the diagonal is also considered.
In theory, standard errors (SEs) approach 0 proportional to $N^{-\frac{1}{2}}$.
In practice, however, each additional participant does not contribute
exactly 1 unit of information.
Relative difference (RD) is a way to transform SEs
into comparable units across conditions,
\[
RD = \frac{SE - SE_{\text{true}}}{SE_{\text{true}}}.
\]
To summarize RDs for a set of parameters,
the $l^2$-norm is used, $||RD||_2$.

The Supplemented EM method admits another opportunity to measure the accuracy of $V$.
The Jacobian (Equation~\ref{eqn:rate-matrix}) is usually not
exactly symmetric and
the final matrix multiplication $\left(I - \RateMat\right)\mathcal{I}(\hat\theta;Y_c)$
may induce further asymmetries.
\textcite{jamshidian2000} pointed out that these asymmetries are pure error
and suggested quantification as the
maximum relative error (MRE) of $V$ with the spectral norm
\begin{align}
\text{MRE}(V) &= ||C^{-\frac{1}{2}} K C^{-\frac{1}{2}}||_2 \label{eqn:mre-v}
\end{align}
where $C=(V+V^T)/2$ is the symmetric part of $V$ and $K=(V-V^T)/2$ is the asymmetric part.
This will be a useful cross-check against our other measures of precision.
After computing $\text{MRE}(V)$, a SEM estimate of $\mathcal{I}(\hat\theta;Y_o)$
is averaged with its transpose to ensure an exactly symmetric matrix.

\subsection{Procedure}

We evaluated convergence properties, accuracy, and elapsed time of
MR-SEM, Tian-SEM, and Agile-SEM with 500 Monte Carlo replications.
The completed data information matrix (Equation~\ref{eqn:info-complete}) and
central difference Richardson extrapolation with an initial step
size of~$10^{-3}$ and 2~iterations
were included as low and high accuracy benchmarks, respectively.
A relative EM tolerance of $10^{-11}$ was used
without EM acceleration.
This relative tolerance roughly corresponds to an absolute tolerance of $10^{-6}$
for the models of interest.
Without EM acceleration,
the EM iteration limit was raised to 750 from the default of 500
to protect many replications of Model cyh1 from early termination.
SEM tolerance was set to the square root of the nominal absolute EM tolerance,
$10^{-\frac{6}{2}}$ \parencite[p.~907]{meng1991}.
Although absolute EM tolerances as low as $10^{-8}$ have been recommended \parencite[p.~318]{cai2008},
such high precision was deemed impractical.
As will be seen,
both MR-SEM and Tian-SEM are already too slow with an absolute EM tolerance of $10^{-6}$.
Raising precision further would make these algorithms even slower.

\begin{table}[h]
\centering
\begin{tabular}{rrrrr}
  \hline
 & RE & Agile & Tian & MR\,SEM \\ 
  \hline
m2pl5 & 2.6 & 3.6 & 3.8 & 4.8 \\ 
  m3pl15 & 1.0 & 1.0 & 1.0 & 1.2 \\ 
  grm20 & 0.0 & 0.4 & 0.0 & 95.4 \\ 
  cyh1 & 0.0 & 0.0 & 20.0 & 70.2 \\ 
   \hline
\end{tabular}
\caption{Percentage of trials that failed to converge by model and algorithm. Failure was due to either iteration limit or a non-positive definite covariance matrix. Since some trials were genuinely unidentified, these trials failed to converge for all algorithms. Compare with the \emph{unidentified} column in Table~\ref{tab:mcData}.} 
\label{tab:posdef-agree}
\end{table}

\begin{table}[h]
\centering
\begin{tabular}{l d{3}d{3}d{3}d{3}d{3} }
\hline
 & RE & Agile & Tian & MR\,SEM & Mstep \\
\hline
m2pl5 \\
\quad  seconds & 0.024 & 0.031 & 0.047 & 0.072 & 0.02 \\
\quad  $\log(D_{KL})$ & 3.225 & 3.232 & 4.537 & 4.364 & 4.532 \\
\quad  $||RD||_2$ & 1.432 & 1.369 & 1.758 & 1.685 & 1.751 \\
m3pl15 \\
\quad  seconds & 0.317 & 0.142 & 0.112 & 0.204 & 0.079 \\
\quad  $\log(D_{KL})$ & 11.893 & 11.893 & 12.014 & 11.944 & 12.014 \\
\quad  $||RD||_2$ & 5.273 & 5.271 & 4.772 & 5.069 & 4.771 \\
grm20 \\
\quad  seconds & 6.834 & 0.451 & 0.466 & 0.226 & 0.032 \\
\quad  $\log(D_{KL})$ & 0.862 & 0.899 & 1.326 & 2.159 & 2.15 \\
\quad  $||RD||_2$ & 0.675 & 0.687 & 0.859 & 1.55 & 1.532 \\
cyh1 \\
\quad  seconds & 38.033 & 2.823 & 9.344 & 11.972 & 0.086 \\
\quad  $\log(D_{KL})$ & 1.406 & 1.395 & 1.482 & 1.626 & 4.919 \\
\quad  $||RD||_2$ & 1.286 & 1.261 & 2.024 & 4.198 & 3.772 \\
\hline \end{tabular}\caption{Mean elapsed time and accuracy of parameter covariance matrix estimators. RE is central difference with Richardson extrapolation and Mstep is the completed data information matrix (Equation~\ref{eqn:info-complete}). Since unconveraged trials were excluded, the performance of MR-SEM and Tian are shown in a most positive light. The scales of $D_{KL}$ and $||RD||_2$ are model specific and should not be compared between models. }
\label{tab:accuracy-speed}

\end{table}

\begin{figure}
\begin{knitrout}
\definecolor{shadecolor}{rgb}{0.969, 0.969, 0.969}\color{fgcolor}
\includegraphics[width=\maxwidth]{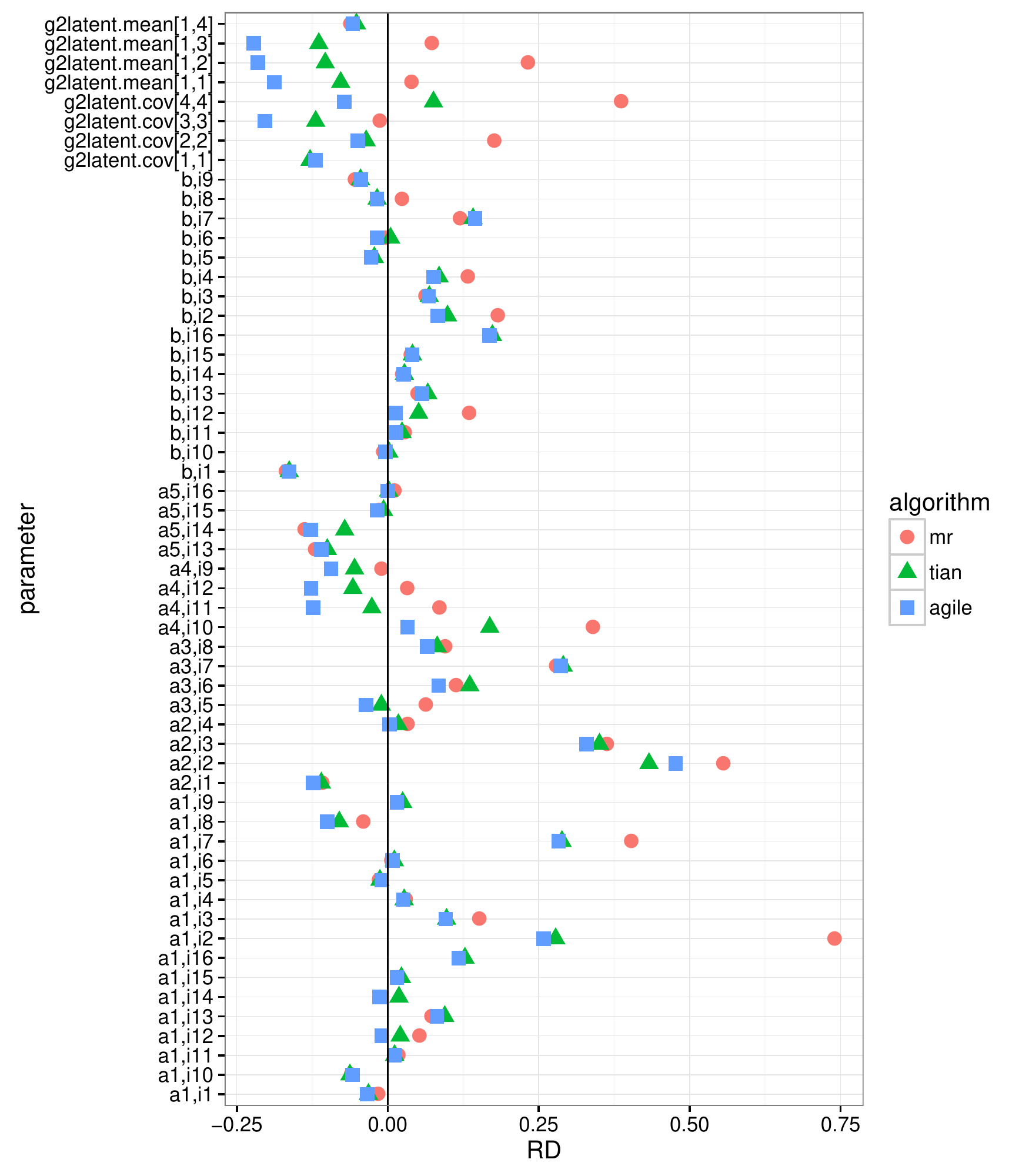} 

\end{knitrout}
\caption{Relative difference of standard errors from the Monte Carlo SEs by algorithm for
a particular replication of Model cyh1. Closer to zero is better.
In this instance, $||RD||_2$ for MR-SEM is 1.38, Agile-SEM is
1, and Tian-SEM slightly outperforms
with 0.94.}
\label{fig:typical-accuracy}
\end{figure}

\section{Results}

Table~\ref{tab:posdef-agree} exhibits the percentage of
models for which each algorithm converged.
MR-SEM failed to
converge for a substantial number of trials where Agile-SEM succeeded.
A failure to converge does not only squander
the time spent due to SEM,
but if SEM is to be reattempted then the model
must be re-fit from starting values.
One of the reasons that Tian-SEM can fail is that sometimes
a parameter arrives at the MLE prior to when Tian-SEM
starts searching the convergence history.
The numerical noise is very intense near the MLE and Tian-SEM has no
ability to move further away.
Another problem applicable to both MR-SEM and Tian-SEM
is that each individual column of the rate matrix
(Equation~\ref{eqn:rij}) is measured at some
random distance from the MLE
(some function of starting values, tolerances, model, and
the amount of data).
This random variability can induce a non-positive definite information matrix.

To provide an intuitive idea of
what the simulation data look like, one replication of Model cyh1
is exhibited in Figure~\ref{fig:typical-accuracy}.
Table~\ref{tab:accuracy-speed} exhibits mean elapsed time and accuracy of
parameter covariance matrix estimators.
Agile-SEM obtained accuracy comparable
to central difference, generally outperforming 
MR-SEM and Tian-SEM even though
Agile's performance was assessed on models for which
these other algorithms did not converge.
We expect $||RD||_2$ and $\log(D_{KL})$ to
be positively correlated.
However, in Model m3pl15, Tian-SEM obtained a better $||RD||_2$
and a worse $\log(D_{KL})$ than Agile-SEM.
For this model, we suggest that Tian-SEM performed similarly
to the M-step standard errors.
Both seem to outperform with respect to $||RD||_2$ but
exhibit relatively poor $\log(D_{KL})$.
The poor $\log(D_{KL})$ suggests that Tian-SEM's
superior $||RD||_2$ performance may not generalize
to different starting values, tolerances, quantities of data, or models.

\section{Discussion and conclusion}

Agile-SEM, a novel variation on Supplemental EM, was developed
with close attention to the limits of
floating-point arithmetic.
Like all Supplemental EM family algorithms,
Agile-SEM evaluates a derivative numerically,
but Agile-SEM carefully controls for the numerical noise intensity
on a per-parameter basis.
We compared the convergence properties, accuracy, and elapsed time of
Supplemental EM family algorithms for a diverse selection IFA models.
Agile-SEM outperformed both MR-SEM and Tian-SEM in all our criteria.

Agile-SEM is a novel method.
More experience is needed with a broad range of IFA models to assess its performance.
Conversely, there is nothing specific to IFA required by Agile-SEM.
It seems likely that Agile could work well on other EM applications.
More work is needed to determine whether the noise model (Equation \ref{eqn:asem-model})
is broadly applicable or specific to item response models.

Although standard errors are a useful tool,
they are not the most accurate way to assess the variability of estimated parameters.
If any parameters are close to a boundary of the feasible set then
likelihood-based confidence intervals should be used instead \parencite[e.g.,][]{pek2015}.
Likelihood-based confidence intervals are comparatively slow to compute,
but offer higher accuracy than a Wald test and are well supported by \OpenMx \parencite{openmx2}.

Complete source code for all algorithms discussed is part of
the \OpenMx source distribution available from \url{http://openmx.psyc.virginia.edu/}.
The \OpenMx website additionally contains documentation
and user support to assist users in
analysis of their own data using item response models and the \textcite{meng1991}
family of algorithms.
Source code for the simulations conducted is available in
the \texttt{inst/models/enormous} subdirectory of the \OpenMx source distribution.
\OpenMx is a package for the \R statistical programming environment \parencite{Rlanguage}.


\printbibliography

\appendix

\section{Item models}\label{app:models}

IFA models involve a set of 
response probability functions to appropriately model the ordinal data.
The response models used in the present article are defined here.
The logistic function,
\[
\text{logistic}(l) \equiv \logit^{-1}(l) \equiv \frac{1}{1+\exp(-l)}
\]
is the basis of the response functions considered here.
Due to the limits of IEEE 754 double-precision binary floating-point,
the maximum absolute logit was set to 35.
That is, $|l| > 35$ was clamped to $|35|$.

\subsection{Dichotomous Model}

The dichotomous response probability can model items when there are exactly two possible
outcomes. It is defined as,
\begin{align*}
\Prob(\pick=0|\bm a,c,g,\bm\tau) &= 1- \Prob(\pick=1|\bm a,c,g,\bm\tau) \\
\Prob(\pick=1|\bm a,c,g,\bm\tau) &= \logit^{-1}(g)+(1-\logit^{-1}(g))\frac{1}{1+\exp(-(\bm a\bm\tau + c))}
\end{align*}
where $\bm a$ is the slope, $c$ is the intercept,
$g$ is the pseudo-guessing lower asymptote expressed in logit units,
and $\bm\tau$ is the latent ability of the examinee \parencite{birnbaum1968}.
A \emph{\#PL} naming shorthand has developed to refer to versions of the dichotomous
model with different numbers of free parameters.
Model $n$PL refers to the model obtained by freeing the
first $n$ of parameters $b$, $a$, and $g$.

\subsection{Graded Response Model}

The graded response model is a response probability function
for 2 or more outcomes \parencite{samejima1969,cai2010b}.
For outcomes k in 0 to K, slope vector $\bm a$, intercept vector $\bm
c$, and latent ability vector $\bm\tau$, it is defined as,
\begin{align*}
\Prob(\pick=0|\bm a, \bm c,\bm\tau) &= 1- \Prob(\pick=1|\bm a,c_1,\bm\tau) \\
\Prob(\pick=k|\bm a, \bm c,\bm\tau) &= \frac{1}{1+\exp(-(\bm a\bm\tau + c_k))} - \frac{1}{1+\exp(-(\bm a\bm\tau + c_{k+1}))} \\
\Prob(\pick=K|\bm a, \bm c,\bm\tau) &= \frac{1}{1+\exp(-(\bm a\bm\tau + c_K))}.
\end{align*}

\subsection{Nominal Model}

The nominal model is a response probability function
for 3 or more outcomes \parencite[e.g.,][]{thissen2010}.
It can be defined as,
\begin{align*}
\bm a &= T_a \bm\alpha \\
\bm c &= T_c \bm\gamma \\
\Prob(\pick=k|\bm s,a_k,c_k,\bm\tau) &= C\ \frac{1}{1+\exp(-(\bm s \bm\tau a_k + c_k))}
\end{align*}
where $a_k$ and $c_k$ are the result of multiplying two vectors
of free parameters $\bm\alpha$ and $\bm\gamma$ by fixed matrices $T_a$ and $T_c$, respectively;
$a_0$ and $c_0$ are fixed to 0 for identification;
and $C$ is a normalizing constant to ensure that $\sum_k \Prob(\pick=k) = 1$.

\section{\proglang{C++} pseudocode implementation}\label{app:pseudocode}

\begin{table}
\centering
\begin{tabular}{l p{10cm} }
  \hline
  Variable & Stores \\
  \hline
\code{Est} & current parameter vector \\
\code{estHistory} & a historical list of parameter vectors \\
\code{freeVars} & count of free parameters \\
\code{maxHistLen} & the maximum number of times that \code{probeEM} could be invoked (integer) \\
\code{offset} & an offset from the parameter's MLE $\hat\theta$ \\
\code{paramProbeCount} & a per-parameter count of calls to \code{probeEM} \\
\code{pick} & the accepted column from \code{rijWork} to copy into \code{rij} \\
\code{probeOffset} & a \code{maxHistLen} by \code{freeVars} matrix of offsets from the MLE (set by \code{probeEM}) \\
\code{rij} & accepted columns from \code{rijWork} \\
\code{rijWork} & a \code{freeVars} by \code{maxHistLen} matrix of candidate Jacobian columns \\
\code{v1} & index of the current parameter into \code{Est} \\
   \hline
\end{tabular}
\caption{Explanation of variables used in the \proglang{C++} pseudocode.}
\label{tab:variables}
\end{table}

In computer code implementation,
Supplemental EM family algorithms do not differ to a great degree.
Two subroutines, \code{probeEM} and \code{recordDiff}, can be profitably factored out.
See Table~\ref{tab:variables} for a description of each variable.%
{\onehalfspacing
\begin{lstlisting}[name=Code]
template <typename T>
void ComputeEM::probeEM(int v1, double offset, /// \label{code:rij}
		        Eigen::MatrixBase<T> &rijWork)
{
  probeOffset(paramProbeCount[v1], v1) = offset;
  Est = optimum;
  Est[v1] += offset;
  // Run EM for a single iteration. Est is updated. /// \label{code:em}
  rijWork.col(paramProbeCount[v1]) = (Est - optimum) / offset;
  paramProbeCount[v1] += 1;
}

template <typename T>
void ComputeEM::recordDiff(int v1, Eigen::MatrixBase<T> &rijWork, /// \label{code:convergence}
			   double *stdDiff, bool *mengOK)
{
  const int h1 = paramProbeCount[v1]-2;
  const int h2 = h1+1;
  Eigen::ArrayXd diff = (rijWork.col(h1) - rijWork.col(h2)).array().abs();
  *mengOK = (diff < semTolerance).all(); /// \label{code:mr-criterion}
  double dist = fabs(probeOffset(h1, v1) - probeOffset(h2, v1));
  *stdDiff = diff.sum() / (diff.size() * dist); /// \label{code:asem-norm}
}
\end{lstlisting}
}%
Function \code{probeEM} at line~\ref{code:rij} implements Equation~\ref{eqn:rij}.
The code is omitted,
but an EM cycle should be run at line~\ref{code:em} (Equation~\ref{eqn:em-map}).
Convergence criteria are checked in function \code{recordDiff} (line~\ref{code:convergence}).
The MR-SEM criterion (Equation~\ref{eqn:mr-criterion}) is implemented on line~\ref{code:mr-criterion}.
The Agile-SEM norm (Equation~\ref{eqn:asem-norm}) is implemented on line~\ref{code:asem-norm}.
The main loop, \code{MengRubinFamily}, iterates over each parameter.
{\onehalfspacing
\begin{lstlisting}[name=Code]
void ComputeEM::MengRubinFamily()
{
  probeOffset.resize(maxHistLen, freeVars);
  paramProbeCount.assign(freeVars, 0);
  Eigen::MatrixXd rij(freeVars, freeVars);

  for (int v1=0; v1 < freeVars; ++v1) {
    Eigen::MatrixXd rijWork(freeVars, maxHistLen);
    int pick = 0;
    bool paramConverged = false;
    if (semMethod == AgileSEM) {
      double offset1 = .001;
      const double stepSize = offset1 * .01;
      probeEM(v1, offset1, rijWork);
      double offset2 = offset1 + stepSize;
      probeEM(v1, offset2, rijWork);
      double diff;
      bool mengOK;
      recordDiff(v1, rijWork, &diff, &mengOK);
      double midOffset = (offset1 + offset2) / 2;
      paramConverged = true;    // always works
      double coef = diff * midOffset * midOffset; /// \label{code:nu}
      offset1 = sqrt(coef/noiseTarget);
      probeEM(v1, offset1, rijWork);
      pick = 2;
    } else if (semMethod == ClassicSEM || semMethod == TianSEM) {
      for (size_t hx=0; hx < estHistory.size(); ++hx) {
	double offset1 = estHistory[hx][v1] - optimum[v1];
        // skip history entries that are too close together
	if (paramProbeCount[v1] &&
        fabs(probeOffset(paramProbeCount[v1]-1, v1) -
        offset1) < tolerance) continue;
        // skip offsets too close to the MLE
	if (fabs(offset1) < tolerance) continue;
	probeEM(v1, offset1, rijWork);
        // at least 2 probes needed to check convergence
	if (hx == 0) continue;
	pick = hx;
	double diff;
	bool mengOK;
	recordDiff(v1, rijWork, &diff, &mengOK);
	if (mengOK) {
	  paramConverged = true;
	  break;
	}
      }
    }

    if (paramConverged) {
      rij.col(v1) = rijWork.col(pick);
    } else {
      return; // failed to converge
    }
  }
  ... /// \label{code:common}
}
\end{lstlisting}
}%
For MR-SEM, \code{estHistory} contains the full EM estimation history whereas
for Tian-SEM, \code{estHistory} only contains parameter vectors near the MLE.
This is the only difference between MR-SEM and Tian-SEM.
Application of the Agile-SEM regression model (Equation~\ref{eqn:asem-model})
is implemented around line~\ref{code:nu}.
The only difference between algorithms is the method to estimate $\RateMat$ (Equation~\ref{eqn:rate-matrix}
stored in variable \code{rij}).
After $\RateMat$ is obtained, the remainder of the algorithm is the same (from line~\ref{code:common} onward).
To facilitate diagnostic output,
the code here stores more data than are strictly needed to complete the computation.

\end{document}